\documentclass[aps,reprint,showpacs,twocolumn,superscriptaddress,floatfix]{revtex4-1}
\usepackage{epsfig}
\usepackage[T1]{fontenc}
\usepackage[utf8]{inputenc}
\usepackage{amsmath}
\usepackage{amssymb}
\usepackage{amsfonts}
\usepackage{mathptmx}
\usepackage{textcomp}
\usepackage{dcolumn}
\usepackage{eucal}
\usepackage{bm}
\usepackage{color}
\usepackage[colorlinks,linkcolor=blue,citecolor=blue]{hyperref}

\usepackage{epstopdf}

\begin{document}

%\bibliographystyle{prsty}%{apsrev}

%%%%%%%%%%%%%%%%%%%%%%%%%%%%%%%%%%%%%%%%%%%%%%%%%%%%%%%%%%%%
%%%%%%%%%%%%%%%%%%%%%%%%%%%%%%%%%%%%%%%%%%%%%%%%%%%%%%%%%%%%
\title{0-$\pi$ phase-controllable \textit{thermal} Josephson junction}

\author{Antonio Fornieri}
\email{antonio.fornieri@sns.it}
\affiliation{NEST, Istituto Nanoscienze-CNR and Scuola Normale Superiore, Piazza S. Silvestro 12, I-56127 Pisa, Italy}

\author{Giuliano Timossi}
%\email{christophe.blanc@nano.cnr.it}
\affiliation{NEST, Istituto Nanoscienze-CNR and Scuola Normale Superiore, Piazza S. Silvestro 12, I-56127 Pisa, Italy}

\author{Pauli Virtanen}
\affiliation{NEST, Istituto Nanoscienze-CNR and Scuola Normale Superiore, Piazza S. Silvestro 12, I-56127 Pisa, Italy}

\author{Paolo Solinas}
\affiliation{SPIN-CNR, Via Dodecaneso 33, I-16146 Genova, Italy}

\author{Francesco Giazotto}
\email{francesco.giazotto@sns.it}
\affiliation{NEST, Istituto Nanoscienze-CNR and Scuola Normale Superiore, Piazza S. Silvestro 12, I-56127 Pisa, Italy}

%%%%%%%%%%%%%%%%%%%%%%%%%%%%%%%%%%%%%%%%%%%%%%%%%%%%%%%%

\date{\today}% It is always \today, today,
             %  but any date may be explicitly specified

%%%%%%%%%%%%%%%%%%%%%%%%%%%%%%%%%%%%%%%%%%%%%%%%%%%%%%%%%%%%
%%%%%%%%%%%%%%%%%%%%%   ABSTRACT         %%%%%%%%%%%%%%%%%%%
%%%%%%%%%%%%%%%%%%%%%%%%%%%%%%%%%%%%%%%%%%%%%%%%%%%%%%%%%%%%

%\begin{abstract}
%\end{abstract}

\pacs{}

%\keywords{Suggested keywords}%Use showkeys class option if keyword
                              %display desired
\maketitle

%\tableofcontents
%%%%%%%%%%%%%%%%%%%%%%%%%%%%%%%%%%%%%%%%%%%%%%%%%%%%%%%%%%%%
%%%%%%%%%%%%%%%%%%%%   INTRODUCTION   %%%%%%%%%%%%%%%%%%%%%%
%%%%%%%%%%%%%%%%%%%%%%%%%%%%%%%%%%%%%%%%%%%%%%%%%%%%%%%%%%%%

\textbf{Two superconductors coupled by a weak link support an equilibrium Josephson electrical current which depends on the phase difference $\varphi$ between the superconducting condensates~\cite{Josephson}. Yet, when a temperature gradient is imposed across the junction, the Josephson effect manifests itself through a coherent component of the heat current that flows oppositely to the thermal gradient for $\vert \varphi \vert <\pi/2$~\cite{MakiGriffin,GiazottoNature,MartinezNature}. The direction of both the Josephson charge and heat currents can be inverted by adding a $\pi$ shift to $\varphi$. In the static electrical case, this effect was obtained in a few systems, e.g. via a ferromagnetic coupling~\cite{Ryazanov,Gingrich} or a non-equilibrium distribution in the weak link~\cite{Baselmans}. These structures opened new possibilities for superconducting quantum logic~\cite{Feofanov,Gingrich} and ultralow power superconducting computers~\cite{Holmes}. Here, we report the first experimental realization of a thermal Josephson junction whose phase bias can be controlled from $0$ to $\pi$. This is obtained thanks to a superconducting quantum interferometer that allows to fully control the direction of the coherent energy transfer through the junction~\cite{FornieriPRB}.
This possibility, joined to the completely superconducting nature of our system, provides temperature modulations with unprecedented amplitude of $\sim$ 100 mK and  transfer coefficients exceeding 1 K per flux quantum at 25 mK. Then, this quantum structure represents a fundamental step towards the realization of caloritronic logic components, such as thermal transistors, switches and memory devices~\cite{MartinezRev,FornieriPRB}. These elements, combined with heat interferometers~\cite{GiazottoNature,MartinezNature,FornieriNature} and diodes~\cite{MartinezNatRect,MartinezAPL}, would complete the thermal conversion of the most important phase-coherent electronic devices and benefit cryogenic microcircuits requiring energy management, such as quantum computing architectures and radiation sensors.}

\begin{figure*}
\centering
%\hspace*{-2.em}
\includegraphics[width=0.7\textwidth]{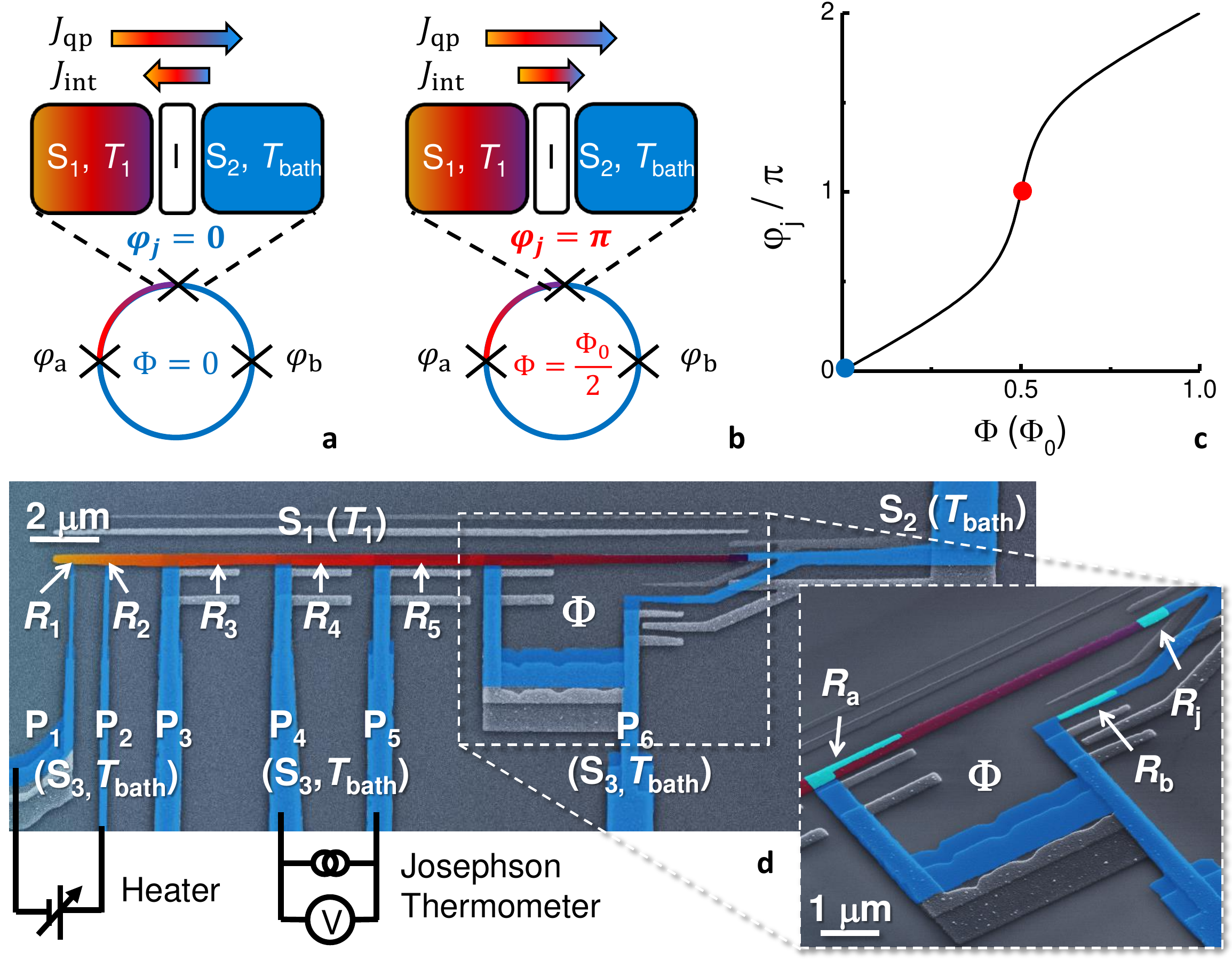}
%\vspace*{-4.ex}
\caption{\textbf{Structure of the $0-\pi$ phase-tunable thermal Josephson junction.} \textbf{a.} Scheme of the temperature-biased tunnel junction "j" between two superconductors S$_1$ and S$_2$ embedded in a three-JJs SQUID. When the flux $\Phi$ threading the superconducting loop is set to 0, the phase difference $\varphi_{\rm j}=0$, and the coherent component of the electronic heat current $J_{\rm int}$ flows oppositely to the thermal gradient. \textbf{b.} When $\Phi=\Phi_0/2$, $\varphi_{\rm j}=\pi$ and $J_{\rm int}$ flows parallel to the other component $J_{\rm qp}$. \textbf{c.} Flux dependence of $\varphi_{\rm j}$ when only a circulating supercurrent is flowing in the interferometer (see text). Blue and red dots represent $\varphi_{\rm j}=0$ and $\varphi_{\rm j}=\pi$, respectively. \textbf{d.} Pseudo-color scanning electron micrograph of the device. The S$_1$ electrode, depicted in a yellow-red gradient, is made of 40-nm-thick Al with $T_{\rm c,1}\simeq 1.3$ K and is coupled to five superconducting probes P$_i$ (with $i=1,2,...5$) and to the lower branch of the SQUID P$_6$ (S$_3$, represented in blue). S$_3$ is composed of a 15-nm-thick Al film with $T_{\rm c,3}\simeq 1.55$ K. On the right side, S$_1$ is connected to the superconductor S$_2$ (also in blue), which consists of a 5-nm-thick Cu and 20-nm-thick Al bilayer with $T_{\rm c,2}\simeq 0.9$ K. P$_1$ and P$_2$ are used as Joule heaters, while P$_3$, P$_4$ and P$_5$ are used to measure the electrical properties of the interferometer and to probe the electronic temperature of S$_1$. The resistances are $R_1\simeq 121$ k$\Omega$, $R_2\simeq 84$ k$\Omega$, $R_3\simeq R_4\simeq 2$ k$\Omega$ and $R_5\simeq 1.5$ k$\Omega$. The blowup is an enlarged image of the SQUID, composed by three JJs with resistances $R_{\rm j}\simeq 3.8$ k$\Omega$, $R_{\rm a}\simeq R_{\rm b}\simeq 2.2$ k$\Omega$. All the junctions present in the structure are implemented through AlO$_{\rm x}$ tunnel barriers, while the area of the loop is about 15 \textmu m$^2$. 
\label{Fig1}}
%% \vspace*{-3.ex}
\end{figure*}

Since the prediction of the Josephson effect~\cite{Josephson}, an insulating barrier connecting two superconductors (an S$_1$IS$_2$ junction) has represented one of the prototypical systems to study macroscopic quantum coherence~\cite{Tinkham}. Nevertheless, its behavior in terms of coherent energy transport has been experimentally investigated only very recently~\cite{GiazottoNature,MartinezNature,MartinezRev,FornieriNature}. It has been shown that if we establish a thermal gradient across a Josephson junction (JJ) by raising the electronic temperature $T_1$ of S$_1$ above the bath temperature $T_{\rm bath}$ (see Fig.~\ref{Fig1}a), a stationary electronic heat current will flow~\cite{MakiGriffin,Guttman,GiazottoAPL}:
\begin{equation}
J_{\rm S_1S_2}(T_1,T_{\rm bath},\varphi)=J_{\rm qp}(T_1,T_{\rm bath})-J_{\rm int}(T_1,T_{\rm bath})\rm\; cos \varphi.\label{Jtot}
\end{equation} 
\noindent Since the Cooper pair condensate carries no entropy under static conditions~\cite{MakiGriffin,Guttman}, $J_{\rm S_1S_2}$ describes the energy transferred by quasiparticles tunneling through the JJ and its direction is determined by the temperature gradient, according to the second principle of thermodynamics. Still, the direction of the second component of $J_{\rm S_1S_2}$ can be arbitrarily regulated by varying $\varphi$. As a matter of fact, $J_{\rm int}$ represents the thermal counterpart of the "quasiparticle-pair interference" contribution to the electrical current that tunnels through a JJ~\cite{Barone,Pop} and its phase coherence allows us to make it flow anti-parallel (if $\varphi=0$) or parallel (if $\varphi=\pi$) to the other component $J_{\rm qp}$, as shown in Figs.~\ref{Fig1}a and~\ref{Fig1}b (see Methods for further details).

%As a matter of fact, $J_{\rm int}$ represents the thermal counterpart of the "quasiparticle-pair interference" contribution to the electrical current that tunnels through a JJ~\cite{Barone,Pop} and its phase coherence allows us to make it interfere destructively (if $\varphi=0$) or constructively (if $\varphi=\pi$). As a consequence, its direction can be parallel or anti-parallel with respect to that of the other component $J_{\rm qp}$, as shown in Figs.~\ref{Fig1}a and~\ref{Fig1}b (see Methods for further details). 

So far, thermal interferometers have been realized to detect and control $J_{\rm int}$, but they could not allow to invert its direction~\cite{GiazottoNature,MartinezNature,FornieriNature}. Here, we demonstrate that a "pseudo" radio frequency superconducting quantum interference device (rf SQUID) composed of three JJs~\cite{FornieriPRB} provides a full control over the coherent component of the heat current exchanged by S$_1$ and S$_2$ through a single JJ, labelled "j". The latter can be polarized from $\varphi_{\rm j}=0$ to $\varphi_{\rm j}=\pi$ by varying the magnetic flux $\Phi$ threading the loop of the SQUID (see Fig.~\ref{Fig1}c). This possibility allows to minimize or maximize $J_{\rm S_1S_2}$, obtaining unprecedented temperature modulation amplitudes and enhanced sensitivities to the magnetic flux. Even more importantly, our structure realizes the fundamental requirement to obtain negative thermal differential conductance which is at the basis of sophisticated non-linear thermal devices, such as tunnel heat diodes~\cite{MartinezAPL}, thermal switches and transistors~\cite{FornieriPRB}.

The implementation of our $0-\pi$ phase-tunable thermal JJ is shown in Fig.~\ref{Fig1}d. The structure was fabricated by electron-beam lithography, three-angle shadow-mask evaporation of metals and \textit{in situ} oxidation (see Methods). It consists of a 40-nm-thick aluminium (Al) island (S$_1$) with a critical temperature $T_{\rm c,1}\simeq 1.3$ K tunnel-coupled to five superconducting probes (P$_{\rm i}$ with $i=1,2,...5$) made of a 15-nm-thick Al film (S$_3$ with $T_{\rm c,3}\simeq 1.55$ K) and acting as Joule heaters or Josephson thermometers~\cite{FornieriPRB,GiazottoRev}. Energy losses from S$_1$ to these probes have been limited by the maximization of the S$_3$ energy gap (see Methods). On the right side, S$_1$ is connected through the tunnel junction "j" to a bilayer composed of 5 nm of copper (Cu) and 20 nm of Al (S$_2$ with a critical temperature $T_{\rm c,2}\simeq 0.9$ K). The latter has been engineered to suppress the Josephson critical current $I_{\rm c,j}$ of junction "j" and to favor the energy transfer from S$_1$ to S$_2$~\cite{FornieriPRB}, which represents the focus of our experiment. Finally, S$_1$ and S$_2$ are connected to a S$_3$ electrode P$_6$ by means of two parallel JJs named "a" and "b". P$_6$, S$_1$ and S$_2$ form a loop with three JJs, two of which are placed on the same branch. This structure has been designed so that $I_{\rm c,j}$ results to be the lowest Josephson critical current in the SQUID, thereby allowing to bias its phase difference from 0 to $\pi$~\cite{FornieriPRB} and change the sign of the second term in Eq.~\ref{Jtot}.

\begin{figure}[h!]
\centering
%\hspace*{-2.em}
\includegraphics[width=0.8\columnwidth]{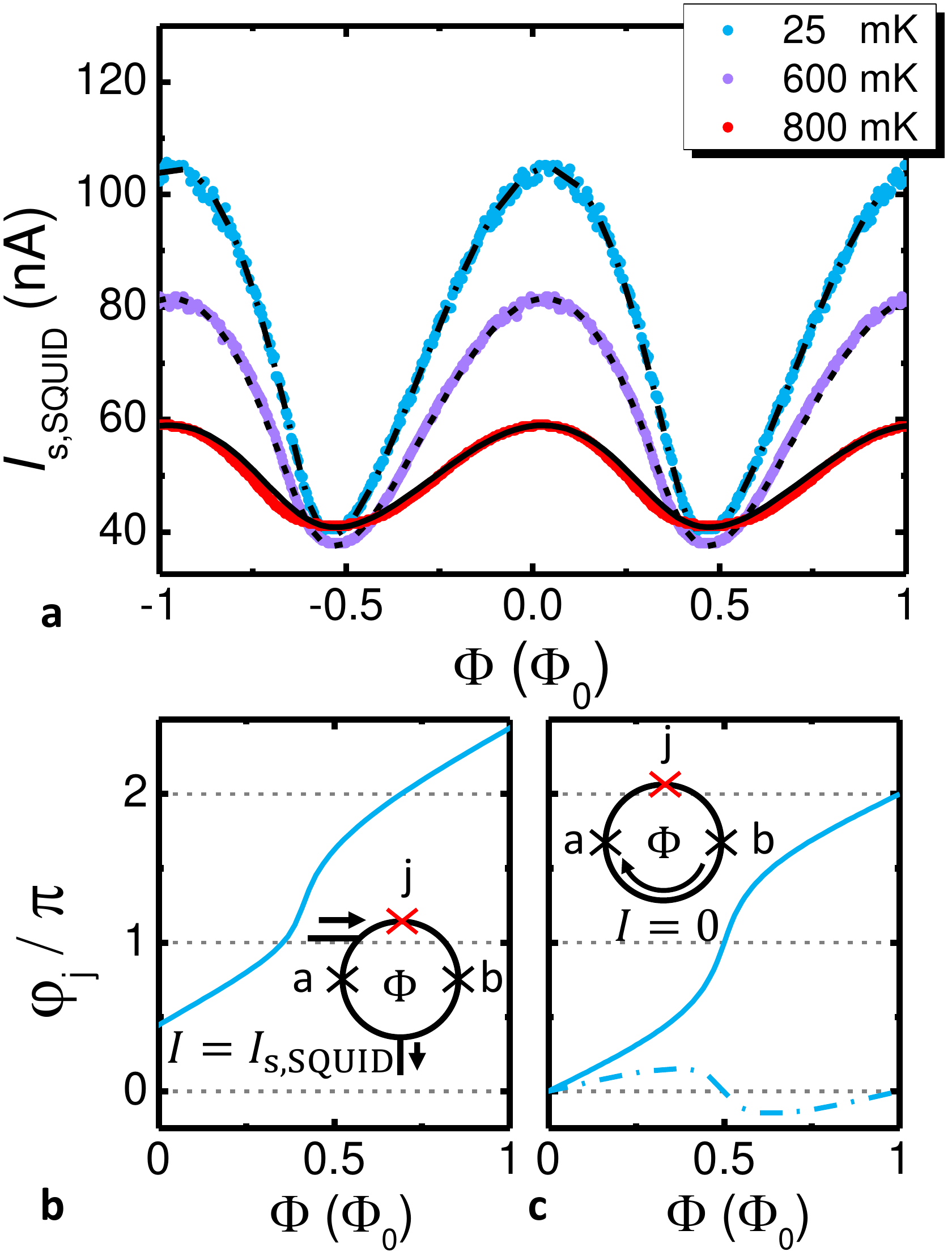}
%\vspace*{-4.ex}
\caption{\textbf{Electrical behavior of the Josephson interferometer.} \textbf{a.} SQUID total switching current $I_{\rm s,SQUID}$ vs. magnetic flux $\Phi$ piercing the loop of the interferometer for selected values of the bath temperature $T_{\rm bath}$. The full circles are experimental values, whereas the black lines are theoretical fits obtained for different values of $I_{\rm a}$, $r_1=I_{\rm a}/I_{\rm j}$ and $r_2=I_{\rm b}/I_{\rm j}$. We extract $I_{\rm a}=72$ nA, $r_1=2.2$ and $r_2=4$ at $T_{\rm bath}=25$ mk, $I_{\rm a}=59.5$ nA, $r_1=2.7$ and $r_2=5$ at $T_{\rm bath}=600$ mK and $I_{\rm a}=49.9$ nA, $r_1=5.5$ and $r_2=6$ for $T_{\rm bath}=800$ mK. \textbf{b.} Calculated behavior of $\varphi_{\rm j}$ vs. $\Phi$ for the supercurrent $I=I_{\rm s,SQUID}$ flowing through the interferometer and for the values of $I_{\rm a}$, $r_1$ and $r_2$ extracted at $T_{\rm bath}=25$ mK. \textbf{c.} Calculated $\varphi_{\rm j}$ vs. $\Phi$ for $I=0$ and for the same values of $I_{\rm a}$, $r_1$ and $r_2$ used in the previous panel. Dash-dotted line represents the calculated $\varphi_{\rm a}$ for the same parameters.
%\textbf{d.} Critical current $I_{\rm c,P_4}$ of the junction between P$_4$ and S$_1$ as a function of the electronic temperature $T_1$. The full circles are the experimental data obtained for $T_1=T_{\rm bath}$, whereas the purple line is the theoretical Ambegaokar-Baratoff expectation (see text for details). Finally, the orange line is the theoretical curve for the case in which $T_1\neq T_{\rm bath}=25$ mK.  
\label{Fig2}}
%% \vspace*{-3.ex}
\end{figure}

The realization of $\pi$ polarization of the junction "j" was verified by first investigating the dissipationless electrical transport through the interferometer via the superconducting electrodes P$_5$ and P$_6$~\cite{Tinkham,Barone}. In order to accurately determine the SQUID critical current $I_{\rm c,SQUID}$, the junction between S$_1$ and P$_5$ was designed to support a larger critical current. We recorded the voltage-current characteristics of the device for different values of $\Phi$, observing a distinct supercurrent branch. The maximum current that can be sustained in this branch is the switching current $I_{\rm s,SQUID}$, which can be substantially different from the expected $I_{\rm c,SQUID}$ because of chemical potential fluctuations that may affect superconducting floating islands~\cite{Tirelli,Quaranta}. In our case, $I_{\rm s,SQUID}$ results to be 50\% smaller than the expected $I_{\rm c,SQUID}$~\cite{Tirelli} (see Methods), but, as we shall argue, this rescaling does not compromise the magnetic interference behavior of the SQUID, which matches very well the theoretical predictions of Ref.~\citenum{FornieriPRB}. Indeed, the value of $I_{\rm s,SQUID}$ is periodically modulated by the magnetic flux piercing the loop, as shown in Fig.~\ref{Fig2}a for selected values of the bath temperature $T_{\rm bath}$. This magnetic-flux interference pattern can be analyzed by first imposing the fluxoid quantization in the loop~\cite{Tinkham} and the Kirchoff laws for the conservation of the supercurrent flowing through the interferometer~\cite{FornieriPRB}. Then, the supercurrent is maximized with respect to $\varphi_{\rm j}$, and $I_{\rm s,SQUID}$ is chosen as the solution that minimizes the Josephson free-energy of the SQUID~\cite{FornieriPRB}.

%In order to exactly determine the SQUID critical current $I_{\rm c,SQUID}$, the junction between S$_1$ and P$_5$ was designed to support a larger critical current. We recorded the voltage-current characteristics of the device for different values of $\Phi$, observing a distinct supercurrent branch. The critical current is the maximum current that can be sustained in this branch, whose value is periodically modulated by the magnetic flux piercing the loop, as shown in Fig.~\ref{Fig2}a for selected values of the bath temperature $T_{\rm bath}$. 

Figure~\ref{Fig2}a displays the very good agreement between the experimental data and the model, which allows us to extract accurate values of the structural parameters of our system. Indeed, for each $T_{\rm bath}$ the theoretical fit provides the switching current $I_{\rm k}$ of each JJ in the SQUID (being $\rm k=a,b,j$) : $I_{\rm j}$ results to be at least 2.2 times smaller than $I_{\rm a}$ and $I_{\rm b}$, guaranteeing an efficient polarization of $\varphi_{\rm j}$. This can be also qualitatively appreciated from the absence of any cusp in the $I_{\rm s,SQUID}$ magnetic interference pattern~\cite{FornieriPRB}. Figure~\ref{Fig2}b shows the flux dependence of $\varphi_{\rm j}$ calculated for the supercurrent $I=I_{\rm s,SQUID}$ flowing through the SQUID and for the values of $r_1=I_{\rm a}/I_{\rm j}$ and $r_2=I_{\rm b}/I_{\rm j}$ obtained from the fit at $T_{\rm bath}=25$ mK. It appears evident how the phase polarization of the junction "j" can span the whole trigonometric period without discontinuities, in contrast to the behavior of a conventional symmetric direct-current SQUID~\cite{SQUIDhandbook,FornieriPRB}. Full control of $\varphi_{\rm j}$ is achieved as well in the configuration used to investigate thermal transport in the device, i.e., in the absence of any Josephson current flowing through the interferometer ($I=0$), and when only a circulating supercurrent is driven by the magnetic flux piercing the loop. This is demonstrated by Fig.~\ref{Fig2}c, which displays the predicted behavior of $\varphi_{\rm j}$ as a function of $\Phi$ in the latter configuration for the structure parameters extracted at $T_{\rm bath}=25$ mK. 

%
%The configuration with only circulating supercurrents in the SQUID is exploited in the thermal measurements which clearly demonstrates how $\varphi_{\rm j}$ can be continuously varied from 0 to 2$\pi$.

\begin{figure*}%[t]
\centering
%\hspace*{-2.em}
\includegraphics[width=0.7\textwidth]{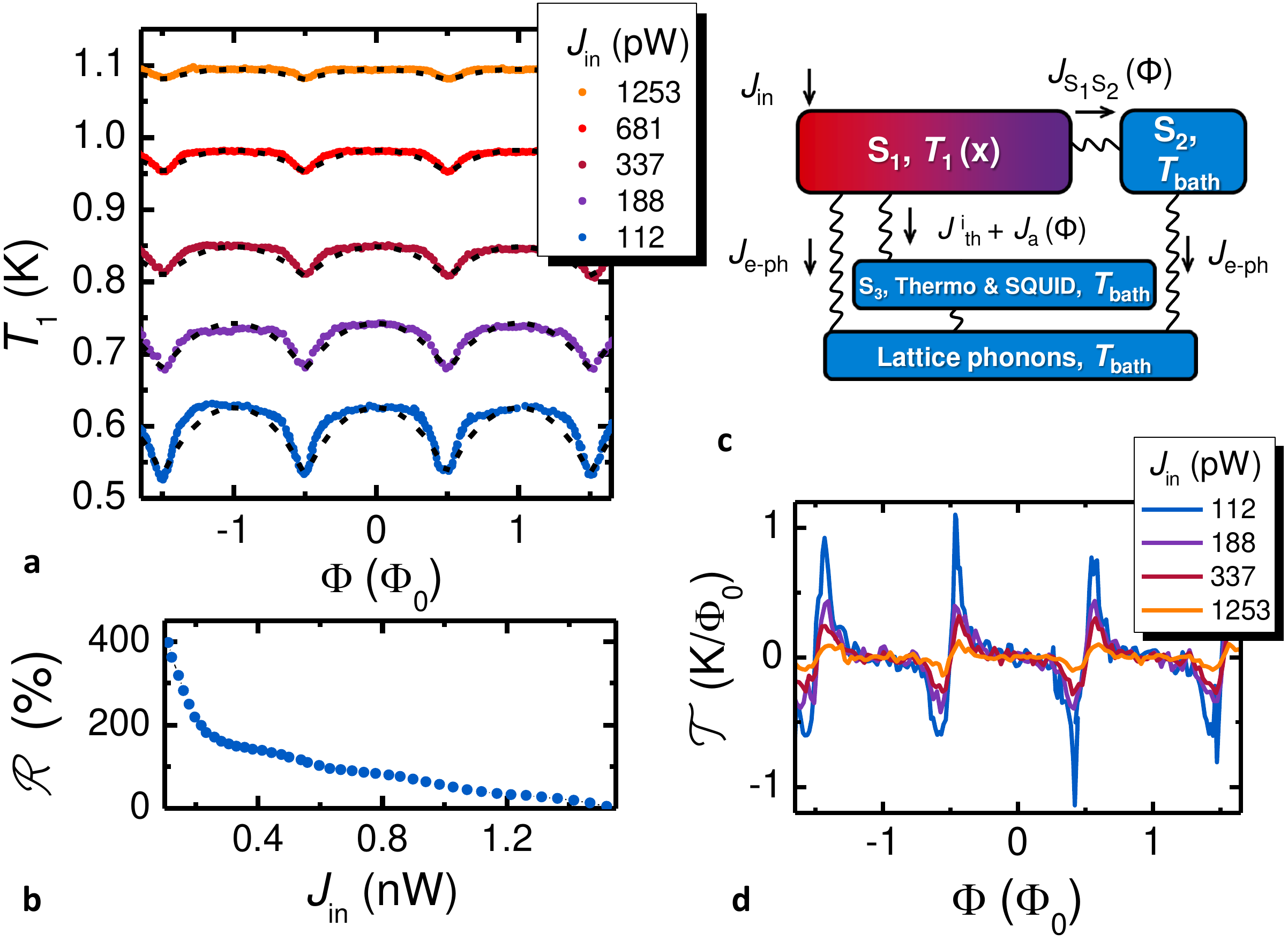}
%\vspace*{-4.ex}
\caption{\textbf{Thermal behavior of the $0-\pi$ phase-controllable Josephson junction at 25 mK.} \textbf{a.} Magnetic flux modulation of S$_1$ temperature ($T_1$) measured by the Josephson thermometer P$_4$ for several values of the injected power $J_{\rm in}$ at $T_{\rm bath}=25$ mK. Filled circles are experimental data, whereas dashed black lines are the theoretical results from the thermal model. \textbf{b.} Relative amplitude modulation $\mathcal{R}=\delta T_1/T_{\rm bath}$ vs. $J_{\rm in}$. \textbf{c.} Thermal model describing the predominant heat exchange processes in our structure. The thermal gradient along S$_1$ is described by the temperature profile $T_1(x)$, while the arrows indicate heat current directions for $T_1(x)> T_{\rm bath}$ (see text). \textbf{d.} Flux-to-temperature transfer function $\mathcal{T}=\partial T_1 /\partial \Phi$ vs. $\Phi$ for selected values of $J_{\rm in}$.\label{Fig3}}
%% \vspace*{-3.ex}
\end{figure*} 
 
The set-up used for thermal measurements is sketched in Fig.~\ref{Fig1}d. The superconducting electrodes P$_1$ and P$_2$ are used to inject Joule power in S$_1$ and raise its electronic temperature above that of the lattice, which we assume to be fully thermalized with the substrate phonons residing at $T_{\rm bath}$, thanks to the negligible Kapitza resistance~\cite{GiazottoNature, MartinezNature,MartinezNatRect,FornieriNature,Wellstood}. In this way, it is possible to generate a thermal gradient across the junctions "j" and "a" of the SQUID, as S$_2$ and P$_6$ are efficiently anchored to $T_{\rm bath}$ owing to their large volumes. This thermal gradient yields finite heat currents $J_{\rm S_1S_2}$ (flowing between S$_1$ and S$_2$ through junction "j") and $J_{\rm a}$ (flowing between S$_1$ and S$_3$ through junction "a"), whose coherent components can be manipulated via the magnetic flux threading the interferometer. This effect leads to a periodic modulation of $T_1$ that can be detected by exploiting the temperature dependence of the Josephson switching current flowing through the junctions between S$_1$ and P$_3$ or P$_4$~\cite{GiazottoRev,FornieriPRB} (see Methods). We note that the variation of $\Phi$ produces only small oscillations of $\varphi_{\rm a}$ around zero (see Fig.~\ref{Fig2}c), so that it will induce rather insignificant corrections to thermal oscillations generated by the $\varphi_{\rm j}$ bias. We also emphasize that the junction "b" will not contribute directly to thermal transport, as it is not temperature biased (see Fig.~\ref{Fig1}d).

We can now focus on the thermal behavior of the structure. Figure~\ref{Fig3}a shows $T_1$ oscillations as a function of the magnetic flux measured by the P$_4$ thermometer for different values of the injected power $J_{\rm in}$ at $25$ mK. Thanks to the experimental design of our structure, these oscillations stem mainly from the modulation of the phase-coherent component of $J_{\rm S_1S_2}$, and the absence of abrupt cusps in the pattern confirms once again the ability of our interferometer to impose $\varphi_{\rm j}=\pi$ for $\Phi=\Phi_0/2$~\cite{FornieriPRB}. As $J_{\rm in}$ increases, the average value of $T_1$ ($\langle T_1\rangle$) raises up to reach almost $T_{\rm c,1}$, while the amplitude of the modulations $\delta T_1$ decreases from a maximum of $\sim 100$ mK and vanishes for $J_{\rm in}\sim 1.6$ nW. The $\pi$ phase polarization, joined to the reduced impact of the electron-phonon coupling in a fully superconducting structure~\cite{Timofeev1}, leads to the largest temperature modulations achieved so far. In particular, up to 400\% of relative modulation amplitude $\mathcal{R}=\delta T_1/T_{\rm bath}$ is obtained for $J_{\rm in}=112$ pW (see Fig.~\ref{Fig3}b), which outscores by more than one order of magnitude the first result obtained in Ref.~\cite{GiazottoNature}.

\begin{figure}%[t]
\centering
%\hspace*{-2.em}
\includegraphics[width=0.85\columnwidth]{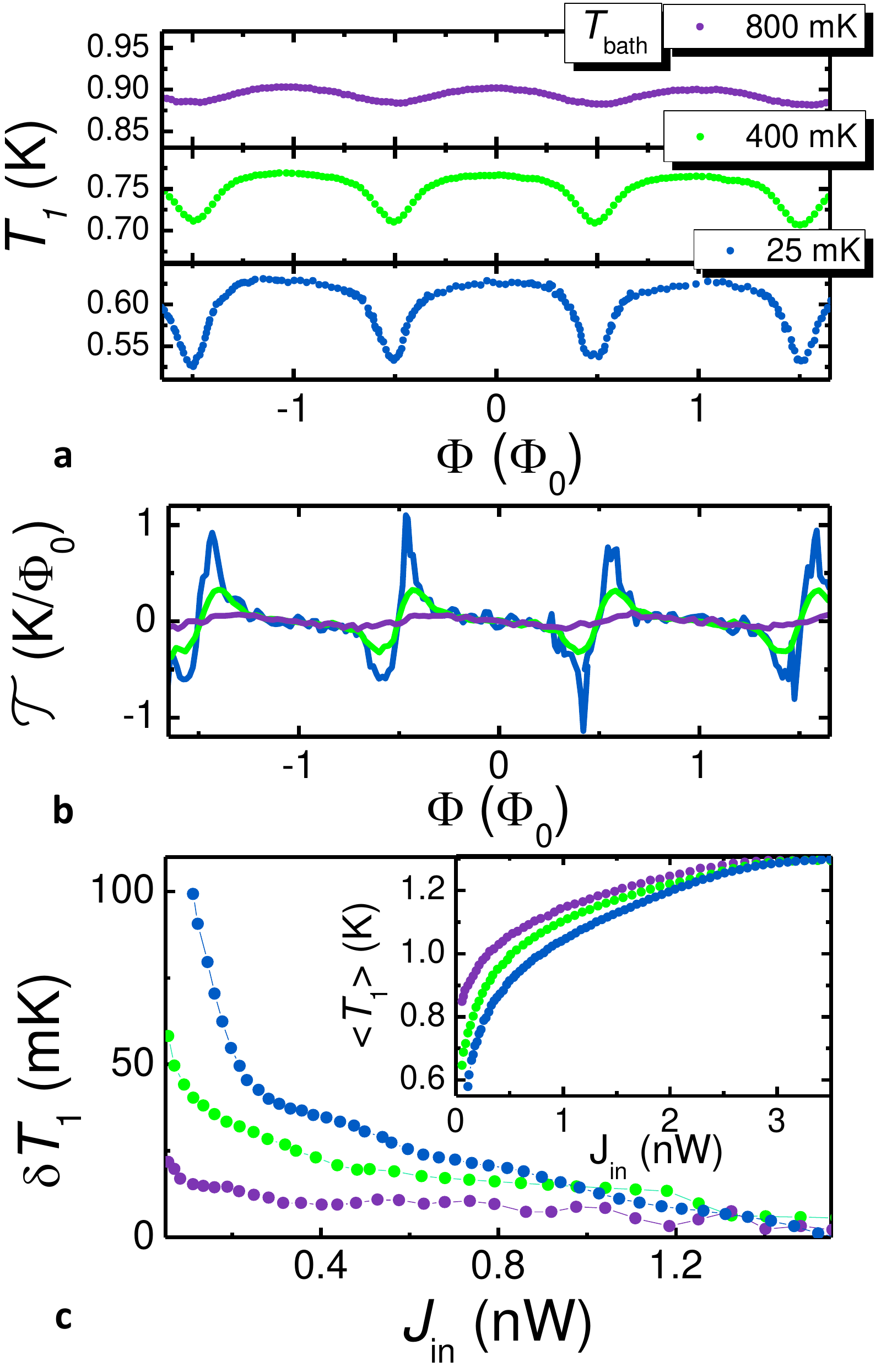}
%\vspace*{-4.ex}
\caption{\textbf{Performance of the $0-\pi$ phase-tunable Josephson junction at different bath temperatures.} \textbf{a.} Magnetic-flux dependent $T_1$ modulations for different values of $T_{\rm bath}$ and for $J_{\rm in}\simeq 112$ pW. \textbf{b.} Transfer function $\mathcal{T}$ as a function of $\Phi$ for the same values of $T_{\rm bath}$ considered in panel a. \textbf{c} Oscillation amplitude $\delta T_1$ and temperature mean value $\langle T_1 \rangle$ (see inset) vs. $J_{\rm in}$ for the same values of $T_{\rm bath}$ shown in the previous panels. In each panel, $T_1$ was measured with the Josephson thermometer P$_4$. \label{Fig4}} 
%% \vspace*{-3.ex}
\end{figure} 

Our observations can be explained by a thermal model (see Fig.~\ref{Fig3}c) outlining all the predominant heat exchange processes present in the structure. Here, terms $J_{\rm th}^{\rm i}$ denote the heat currents delivered by S$_1$ to the thermometer electrodes P$_{\rm i}$, being $\rm i=3,4,5$, and $J_{\rm e-ph}$ describes the power released by S$_1$ to lattice phonons residing at $T_{\rm bath}$ (see Methods for details). Temperature measurements obtained from different thermometers indicate the presence of a thermal gradient along the S$_1$ electrode for $\langle T_1 \rangle \gtrsim 0.7$ K (see Methods). Therefore, we model S$_1$ as a one-dimensional diffusive superconductor with a temperature profile $T_1(x)$, where $x$ is the coordinate along the electrode. $T_1(x)$ can be obtained from the stationary heat diffusion equation~\cite{GiazottoRev,Timofeev2}:
\begin{equation}
\frac{d}{dx}\left\lbrace \kappa[T_1(x)] \frac{dT_1(x)}{dx}\right\rbrace=\frac{J_{\rm e-ph}(x)}{\mathcal{V}_1}+\frac{J_{\rm a}(x)}{\mathcal{V}_{\rm a}}+\sum_{i=3,4,5} \frac{J_{\rm th}^{\rm i}(x)}{\mathcal{V}_{\rm th}^{\rm i}}, \label{heateq}
\end{equation}
where $\kappa[T_1(x)]$ and $\mathcal{V}_1$ are the electronic heat conductivity~\cite{Timofeev2} and the volume of S$_1$, respectively. Moreover, $\mathcal{V}_{\rm th}^{\rm i}=l_{\rm i}\, A$ and $\mathcal{V}_{\rm a}=l_{\rm a}\, A$, being $A$ the cross-section of S$_1$, $l_{\rm i}$ the length of the $i$-th thermometer junction (with $i=3,4,5$) and $l_{\rm a}$ the length of the SQUID junction "a". As boundary conditions, we impose $J_{\rm in} = -\kappa[T_1(0)]AT'_1(0)$ and $J_{\rm S_1S_2} = -\kappa[T_1(l)]AT'_1(l)$, determining the relation between the temperature gradients $T'_1(x)$ and the heat fluxes at the ends of S$_1$.
%\begin{align} 
%&J_{\rm in} = \left. -\kappa[T_1(0)]A\frac{dT_1(x)}{dx}\right|_{x=0}, \\ 
%&J_{\rm S_1S_2} =\left. -\kappa[T_1(l)]A\frac{dT_1(x)}{dx}\right|_{x=l}.
%\end{align}
Here, the junction between S$_1$ and P$_2$ is set to $x=0$, while the junction "j" of the SQUID is set to $x=l=17.5$ \textmu m (see more details in the Methods).
The model accounts for the thermal budget in S$_1$, and neglects photon-mediated thermal transport owing to poor impedance matching among the different electrodes of the structure~\cite{Pascal,meschke,Bosisio}.

The calculated $T_1(l_4)$ (being $l_4$ the position of the P$_4$ junction) was fitted to the temperature measured with the thermometer P$_4$ by using the structure parameters determined from the electrical measurements, and by varying $J_{\rm in}$ and $R_{\rm j}$ as fitting parameters (see Methods for further details). The model provides a good agreement with the data (see Fig.~\ref{Fig3}a), accounting for the predominant heat transport mechanisms in our system, and confirming the $0-\pi$ tunability of the junction "j". 

Figure~\ref{Fig3}d displays the flux-to-temperature transfer coefficient $\mathcal{T}=\partial T_1 /\partial \Phi$ for different values of $J_{\rm in}$. Since our interferometer is not a standard rf SQUID, $\varphi_{\rm j}$ presents a steep increase in the variation from $\pi/2$ to $3\pi/2$, as shown in Fig.~\ref{Fig2}c. This is reflected by the sharp minimum of the thermal oscillations, which exhibit a maximum $\vert \mathcal{T}\vert$ exceeding 1 K/$\Phi_0$. The latter corresponds to a sensitivity five times larger than that previously achieved~\cite{GiazottoNature,MartinezNature,FornieriNature}.

The impact of bath temperature on our $0-\pi$ phase-tunable JJ is illustrated in Fig.~\ref{Fig4}. Figure~\ref{Fig4}a displays $T_1(\Phi)$ at $J_{\rm in}\simeq 112$ pW for different values of $T_{\rm bath}$. The amplitude of the oscillations and the transfer coefficient (see Fig.~\ref{Fig4}b) are progressively suppressed by the temperature-driven enhancement of the electron-phonon coupling, although the effect is evidently softer with respect to that occurring in normal metal electrodes~\cite{Wellstood,GiazottoNature,MartinezNature,MartinezNatRect,FornieriNature}. As a matter of fact, $\delta T_1$ remains as large as 20 mK at $T_{\rm bath}=800$ mK, which almost doubles the maximum operation temperature achieved so far~\cite{GiazottoNature,MartinezNature,MartinezNatRect}. 
Finally, Fig.~\ref{Fig4}c summarizes the overall behavior of our system. While $\delta T_1$ vanishes for increasing values of $J_{\rm in}$, the mean value $\langle T_1 \rangle$ tends towards saturation for $J_{\rm in}>2$ nW due to the stronger impact of the electron-phonon coupling for $T_1 \sim T_{\rm c,1}$.

In summary, we have realized the first $0-\pi$ phase-controllable thermal JJ able to invert the sign of the coherent component of the heat current exchanged by two superconductors. This result has been achieved via a "pseudo" rf SQUID formed by three JJs. The full tunability of the $0-\pi$ thermal JJ, joined to the fully superconducting nature of our device, leads to temperature oscillations with amplitudes as high as 100 mK and flux-to-temperature transfer coefficients exceeding 1 K$/\Phi_0$ at 25 mK. The system can work up to 800 mK of bath temperature, and represents a crucial step towards the design of more exotic caloritronic devices. For instance, the $\pi$ phase-bias would enable the observation of the negative differential thermal conductance~\cite{FornieriPRB}, which is a necessary requisite to get thermal hysteresis and heat amplification~\cite{FornieriPRB}. These effects are of strong impact and relevance for the realization of solid-state thermal memories and heat transistors~\cite{FornieriPRB}, breaking ground for the conception of thermal logic gates~\cite{LiRev} and advanced phase-coherent caloritronic circuits.

\section*{Acknowledgments}
We acknowledge the MIUR-FIRB2013–Project Coca (grant no. RBFR1379UX), the European Research Council under the European Union’s Seventh Framework Programme (FP7/2007-2013)/ERC grant agreement no. 615187 - COMANCHE and the European Union (FP7/2007-2013)/REA grant agreement no. 630925 – COHEAT for partial financial support.

\section*{Author contributions}
A.F. fabricated the samples. A.F. and G.T. performed the measurements. A.F. and G.T. analysed the data and carried out the simulations with inputs from P.V., P.S. and F.G. A.F. and F.G. conceived the experiment. All the authors discussed the results and their implications equally at all stages, and wrote the manuscript.

\begin{widetext}
\section*{Methods}
\setcounter{figure}{0}
\renewcommand{\figurename}{EXTENDED DATA FIG.}

\subsection*{Sample fabrication}
The devices were fabricated with electron-beam lithography and three-angle shadow-mask evaporation of metals onto an oxidized Si wafer through a bilayer resist mask. The evaporation and oxidation were made using an ultra-high vacuum electron-beam evaporator, which allowed us to deposit first 15 nm of Al at an angle of 40$^{\circ}$ to form the superconducting probes P$_i$, with $\rm i=1,...,6$. Then the sample was exposed to 100 mTorr of O$_{2}$ for 5 minutes to realize the thin insulating layer of AlO$_{\rm x}$ forming the tunnel-barriers in all the P$_i$ junctions and in the junction "b" of the SQUID. Afterwards, the sample was tilted at an angle of 30$^{\circ}$ and a deposition of 5 nm of Cu and 20 nm of Al was performed to implement the superconducting lead S$_2$. Another exposition of 100 mTorr of O$_{2}$ for 5 minutes was required to realize the insulating layers of the junctions "a" and "j" forming the SQUID. Finally, 40 nm of Al were evaporated at 0$^{\circ}$ to deposit the superconducting island S$_1$. The S$_1$ electrode has a volume $\mathcal{V}_1=2.7\times 10^{-19}$ m$^3$, while the other electrodes have a volume $\mathcal{V}\sim 10^{-15}$ m$^3$.

\subsection*{Measurements}
All the measurements have been performed in a filtered dilution refrigerator down to 25 mK. The Josephson thermometers were current biased by means of a low-noise floating source, while the voltage drop across the junctions was monitored with a standard room-temperature preamplifier. All the values of the temperature $T_1$ were extracted from an average of at least 20 switching current measurements. On the other hand, the heaters were piloted with voltage biasing in the range 0-40 mV, corresponding to a maximum Joule power of $\sim 4$ nW injected in the S$_1$ electrode.

\subsection*{Characterization of the device by means of quasiparticle tunneling}
In order to fully characterize our system, we investigated quasiparticle transport through all the tunnel junctions present in the structure. First, we measured the current-voltage characteristic of the junction connecting S$_1$ to the probe P$_2$ (junction 2) to extract the parameters of superconductors S$_1$ and S$_3$ (see Fig~\ref{Fig1}d). Towards this end, we applied a voltage bias $V$ to the electrodes P$_2$ and P$_3$ and measured the current flowing through the series of junctions 2 and 3 (the latter connecting S$_1$ and P$_3$). Because of the large asymmetry in the junction parameters, we observed the quasiparticle current of only the small junction 2, while the larger junction 3 remained in the supercurrent state~\cite{Timofeev1,Tirelli}. If S$_1$ resides at the electronic temperature $T_1$ and the superconductor S$_3$ at $T_3$, the stationary quasiparticle current flowing through the tunnel junction is given by~\cite{Tinkham,Barone,GiazottoRev}:
\begin{equation}
I_{\rm qp}=\frac{1}{R_2e}\int_{-\infty}^{\infty} \mathrm{d}\epsilon \mathcal{N}_1(\epsilon-eV, T_1)\mathcal{N}_3(\epsilon , T_3)[f(\epsilon-eV, T_1)-f(\epsilon, T_3)],
\label{Iqp}
\end{equation}
where $\mathcal{N}_{1,3}(\epsilon, T_{\rm 1,3})=|\Re[(\epsilon+i\Gamma_{1,3})/\sqrt{(\epsilon+i\Gamma_{1,3})^2-\Delta_{1,3}^2(T_{\rm 1,3})}]|$ are the smeared normalized Bardeen-Cooper-Schrieffer densities of states (BCS DOSs) in the superconductors S$_1$ and S$_3$~\cite{Tinkham}, $f(\epsilon ,T_{1,3})=[1+\text{exp}(\epsilon/k_{\rm B} T_{1,3})]^{-1}$ are the Fermi-Dirac distributions, $\Gamma_{1,3}=\gamma_{1,3} \Delta_{1,3}(0)$ are the Dynes parameters accounting for the quasiparticle finite lifetime~\cite{Dynes}, $\Delta_{1,3}(T_{1,3})$ are the temperature-dependent energy gaps, $R_{\rm 2}$ is the tunnel junction normal-state resistance, $e$ is the electron charge and $k_{\rm B}$ is the Boltzmann constant.

\begin{figure*}[h!]
\centering
%\hspace*{-2.em}
\includegraphics[width=0.6\textwidth]{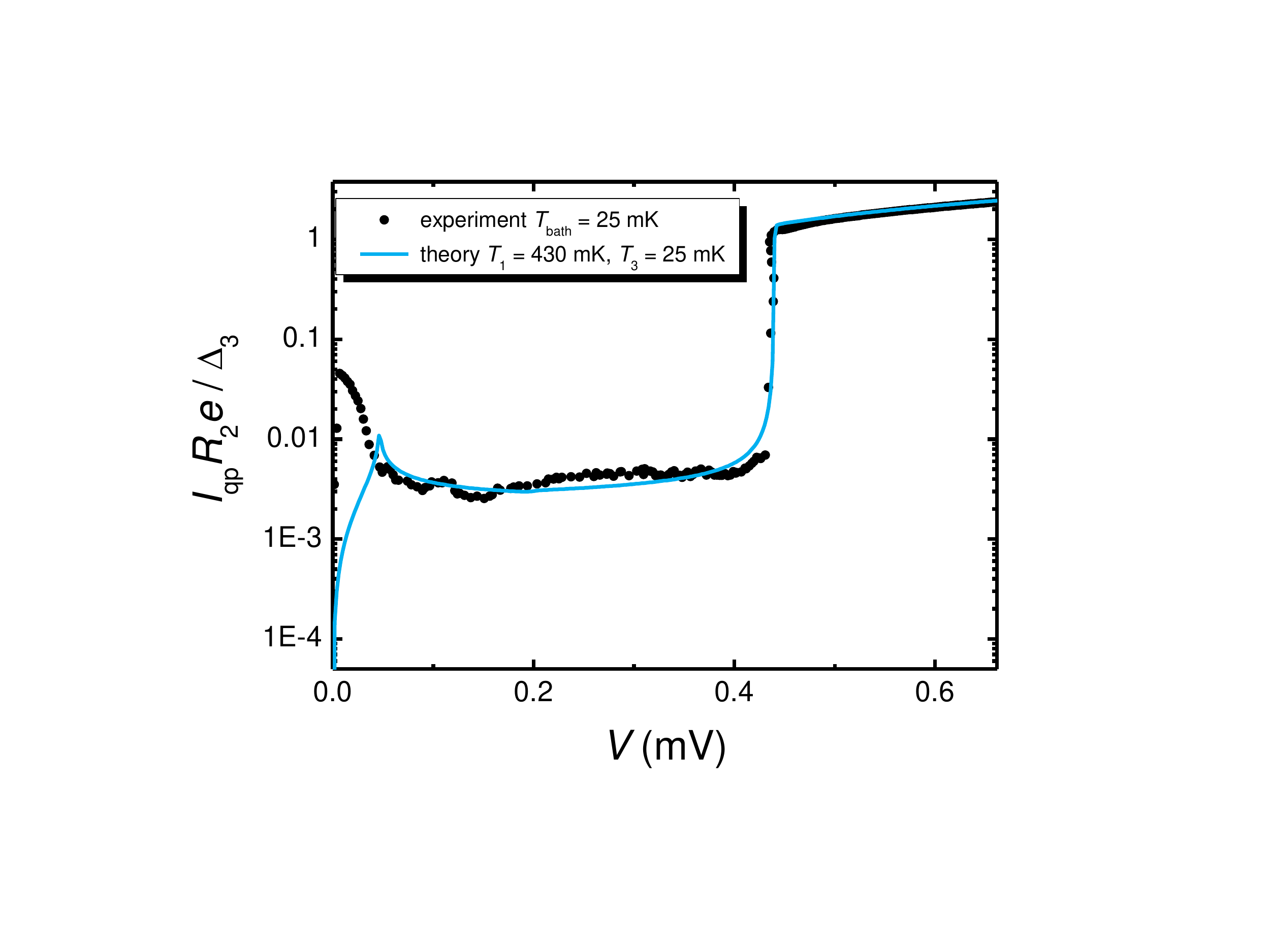}
%\vspace*{-4.ex}
\caption{\textbf{Quasiparticle measurement of the electronic temperature of S$_1$.} Normalized current-voltage characteristic of the tunnel junction connecting S$_1$ to P$_2$ at a bath temperature of 25 mK. The full circles are experimental data while the blue solid line is the theoretical fit (not taking into account the Josephson current at $V=0$). The fit was obtained by setting $\Delta_1\simeq 200\;\mu$eV, $\Delta_3=240\; \mu$eV, $R_2=84$ k$\Omega$ and $\gamma_1=\gamma_3=5\times 10^{-4}$. The extracted electronic temperatures are $T_1\simeq 430$ mK and $T_3\simeq 25$ mK.\label{ExtFig1}} 
%% \vspace*{-3.ex}
\end{figure*} 

Extended Data Figure~\ref{ExtFig1} shows the experimental data measured at $T_{\rm bath}=25$ mK along with the theoretical fit obtained from Eq.~\ref{Iqp}. The model does not take into account the experimental peak around $V=0$ generated by the Josephson effect, but allows to extract relevant structural parameters, such as $\Delta_1(0)\simeq 200\; \mu$eV, $\Delta_3(0)\simeq 240\; \mu$eV, $\gamma_{1}=\gamma_3=5\times10^{-4}$, $T_3=T_{\rm bath}=25$ mK and $T_1\simeq 430$ mK. The difference between the electronic temperatures of S$_1$ and P$_2$ may be ascribed to the large difference in the volumes of the electrodes (P$_2$ extends to a large pad with a volume four order of magnitudes larger than $\mathcal{V}_1$), which reflects on the electron-phonon coupling acting as an important channel of energy release in superconductors. This result is consistent with previous experiments on the energy relaxation of superconducting floating islands~\cite{Timofeev1} and does not affect our experiment, since all the results are obtained for $T_1>500$ mK. Moreover, we note that the Dynes parameters $\gamma_{1,3}$ increase of about two orders of magnitude as $T_{\rm bath}$ is raised up to 1.2 K. A possible explanation for this behavior is the environmental photon-assisted tunneling~\cite{PekolaPRL2010}.

To further analyze our device, we measured the current-voltage characteristics of the "pseudo" rf SQUID via the electrodes P$_6$ and S$_2$ at different bath temperatures. The quasiparticle current exhibits two well-defined singularity-matching peaks~\cite{Barone,Tinkham} at the voltages $V_{1-2}=\Delta_1(T_{\rm bath})-\Delta_2(T_{\rm bath})$ and $V_{3-2}=\Delta_3(T_{\rm bath})-\Delta_2(T_{\rm bath})$ originating from the parallel junctions "j" and "b". These values, together with the voltage $V_{\rm 1+2}=\Delta_1(T_{\rm bath})+\Delta_2(T_{\rm bath})$, allow to extract the temperature dependence of the superconducting energy gaps $\Delta_{\rm 1,2,3}$, as shown in Extended Data Fig.~\ref{ExtFig2}. The experimental data (full circles) are very well described by the Bardeen-Cooper-Schrieffer temperature dependence of the superconducting gap~\cite{Tinkham}, confirming the ideal electric behavior of our system. In summary, we obtain $T_{\rm c,1}\simeq 1.3$ K, $T_{\rm c,2}\simeq 0.9$ K, $T_{\rm c,3}\simeq 1.55$ K, $\Delta_1(0)\simeq 200\; \mu$eV, $\Delta_2(0)\simeq 120\; \mu$eV and $\Delta_3(0)\simeq 240\; \mu$eV. These values reflect the different thickness and composition of the superconducting leads forming our structure, and are essential to limit the energy release from S$_1$ to S$_3$ while favoring the heat exchanged between S$_1$ and S$_2$ through the junction "j" of the SQUID. Finally, the electrical characterization allows us to measure the normal-state resistance of each junction present in the device: $R_1\simeq 121$ k$\Omega$, $R_2\simeq 84$ k$\Omega$, $R_3\simeq R_4\simeq 2$ k$\Omega$, $R_5\simeq 1.5$ k$\Omega$, $R_{\rm j}\simeq 3.8$ k$\Omega$ and $R_{\rm a}\simeq R_{\rm b}\simeq 2.2$ k$\Omega$.

\begin{figure*}[h!]
\centering
%\hspace*{-2.em}
\includegraphics[width=0.6\textwidth]{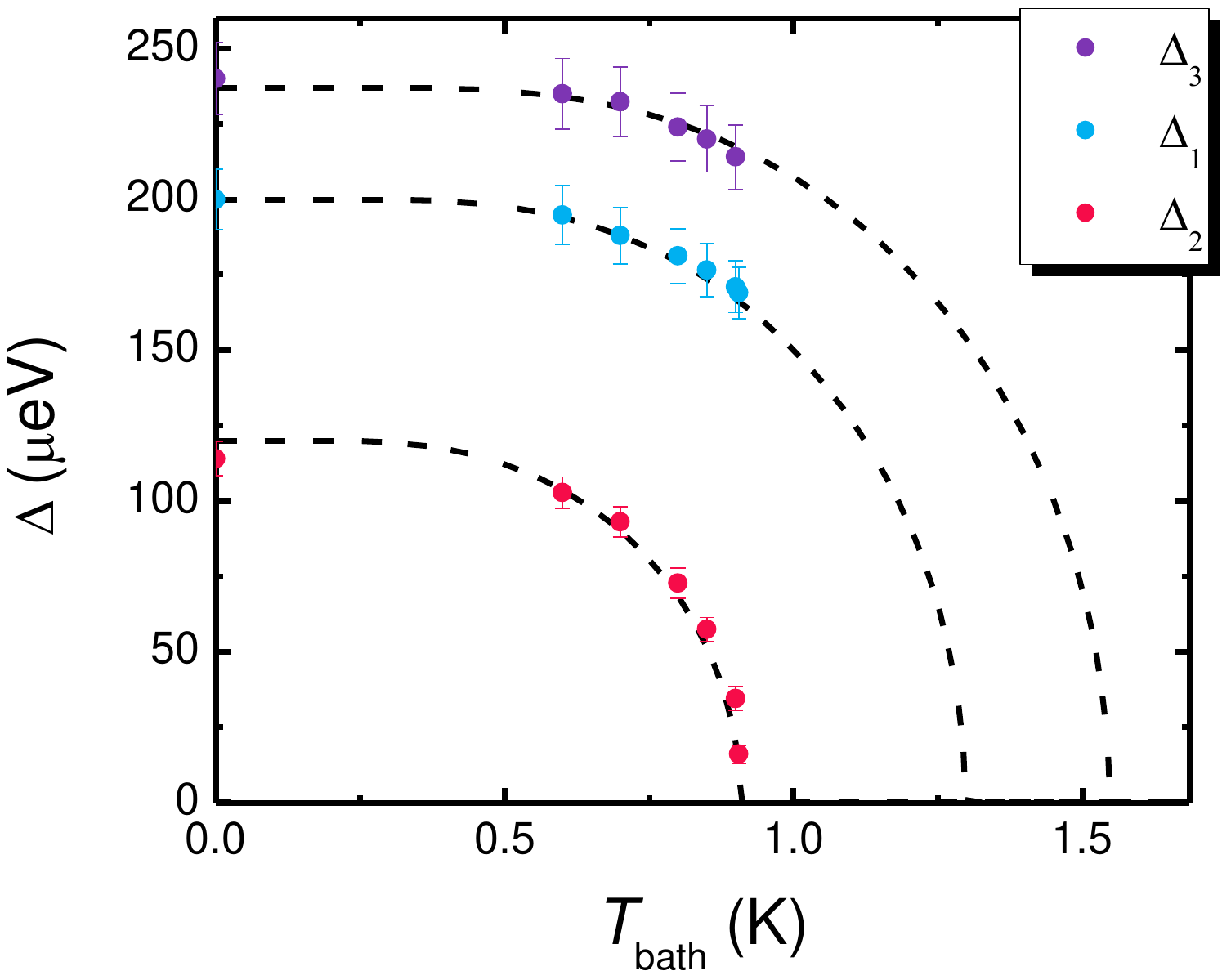}
%\vspace*{-4.ex}
\caption{\textbf{Temperature dependence of the superconducting energy gaps.} Energy gaps $\Delta_{1,2,3}$ vs. $T_{\rm bath}$. The full circles are experimental data extracted from the position of the singularity-matching peaks in current-voltage characteristics of the SQUID (see Methods). The dashed lines are the Bardeen-Cooper-Schrieffer predictions for the temperature dependence of the superconducting energy gaps.\label{ExtFig2}} 
%% \vspace*{-3.ex}
\end{figure*}

\subsection*{Josephson current and thermometer calibration}
In the thermal measurement set-up, the electronic temperature $T_1$ of S$_1$ was measured by exploiting the temperature dependence of the Josephson switching currents $I_{\rm s,3}$ and $I_{\rm s,4}$ flowing through the JJs formed by S$_1$ with P$_3$ and P$_4$, respectively (see Fig~\ref{Fig1}d). The calibration of the Josephson thermometer was obtained by measuring $I_{\rm s,4}$ in equilibrium at several values of the bath temperature $T_{\rm bath}$, as shown in Extended Data Fig.~\ref{ExtFig3}. The experimental data can be successfully fitted by the generalized Ambegaokar-Baratoff formula~\cite{FornieriPRB,GiazottoJap,Tirelli} for the critical current of a JJ between the superconductors S$_1$ and S$_3$ residing at the electronic temperatures $T_1$ and $T_3$, respectively:
\begin{equation}
I_{\rm c}(T_1,T_3)=\frac{1}{2eR_{\rm T}}\left|\displaystyle\int_{-\infty}^{\infty} \mathrm{d}\epsilon\lbrace \textbf{f}(\epsilon,T_1)\Re[\mathcal{F}_1(\epsilon,T_1)]\Im[\mathcal{F}_3(\epsilon,T_3)]
+\textbf{f}(\epsilon,T_3)\Re[\mathcal{F}_3(\epsilon,T_3)]\Im[\mathcal{F}_1(\epsilon,T_1)]\vphantom{\frac{1}{2}}\right|.
\end{equation}
Here, $\textbf{f}(\epsilon,T_{1,3})=\mathrm{tanh}(\epsilon/2k_{\rm B}T_{1,3})$, $\mathcal{F}_{1,3}(\epsilon,T_{1,3})=\Delta_{1,3}(T_{1,3})/\sqrt{(\epsilon+i\Gamma_{1,3})^2-\Delta_{1,3}^2(T_{1,3})}$ are the anomalous Green's functions in the superconductors~\cite{Barone} and $R_{\rm T}$ is the junction normal-state resistance. The model was fitted to experimental data by using the parameters extracted from the electrical characterization of the device, while we set $R_{\rm T}=4.58$ k$\Omega$ as the only fitting parameter. A possible reason for $R_{\rm T}$ being higher than the nominal value of $R_4\simeq 2$ k$\Omega$ may be ascribed to fluctuations of the chemical potential in the superconducting island S$_1$, which can induce time-dependent phase evolutions, thereby reducing the maximum supercurrent supported by the JJ~\cite{Tirelli}, as also observed for the SQUID JJs (see Main Text). 

The good agreement between the experiment and the fit allows to extract the thermometer calibration for the configuration in which $T_1>T_{\rm bath}$ (see Extended Data Fig.~\ref{ExtFig3}). For each value of the injected power $J_{\rm in}$ and the magnetic flux $\Phi$ piercing the SQUID, this calibration is used to convert the measured $I_{\rm s,4}$ into the temperature $T_1$.

\begin{figure*}[h!]
\centering
%\hspace*{-2.em}
\includegraphics[width=0.6\textwidth]{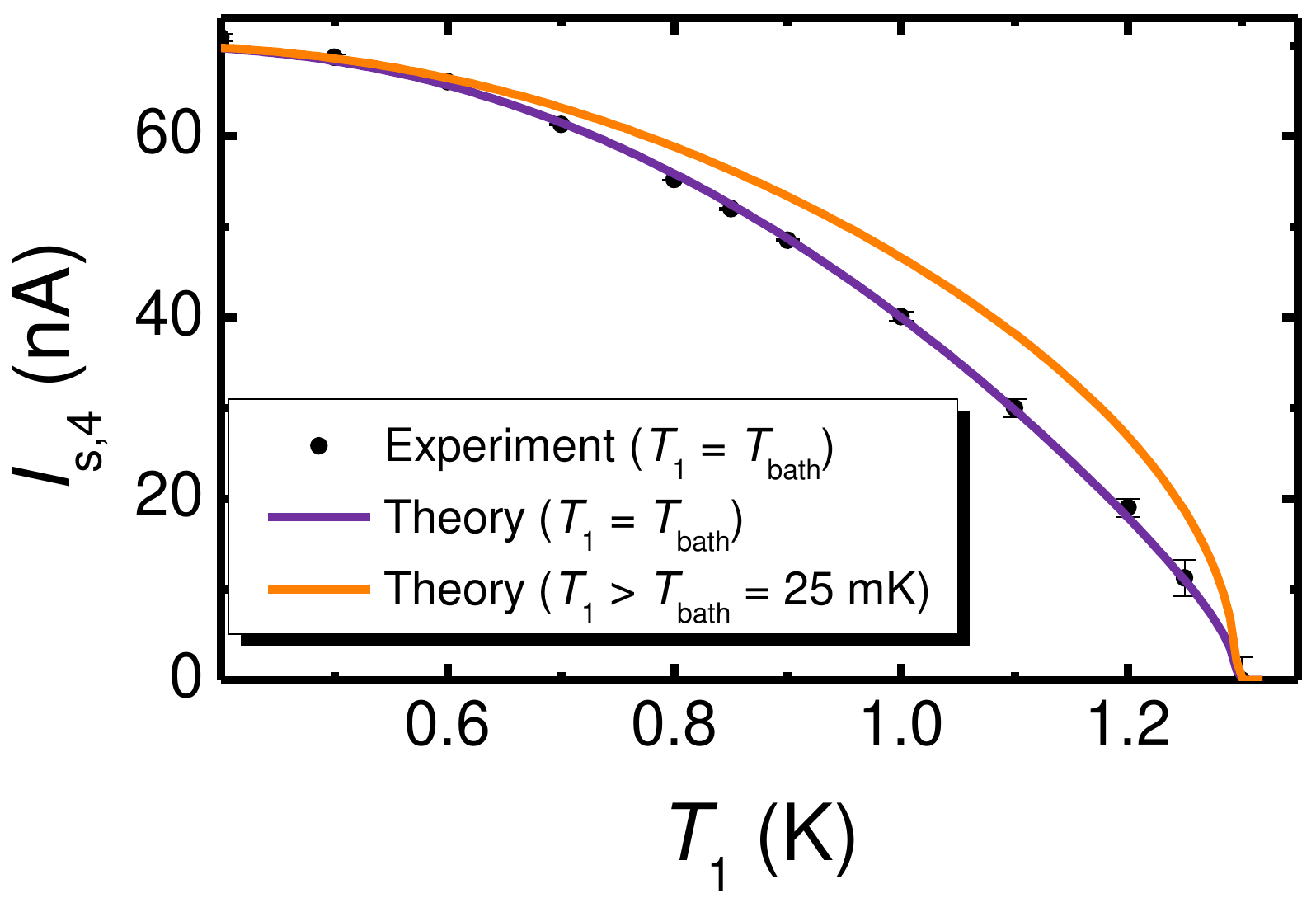}
%\vspace*{-4.ex}
\caption{\textbf{Calibration of a Josephson thermometer.} Switching current of the JJ between S$_1$ and P$_4$ vs. $T_1$. The full circles are experimental data measured in the configuration $T_1=T_{\rm bath}$. The purple solid line is the theoretical fit based on the generalized Ambegoakar-Baratoff prediction (see Methods), while the orange line is the expectation for the configuration $T_1>T_{\rm bath}=25$ mK.\label{ExtFig3}} 
%% \vspace*{-3.ex}
\end{figure*}

\subsection*{Temperature gradient in S$_1$ and thermal model}
Since S$_1$ is a 20-\textmu m-long superconducting thin film, its electronic temperature may become non-uniform as we increase the Joule power injected by the heaters. This hypothesis is confirmed by the measurement of $T_1$ with two thermometers placed at different distances from the heater junctions. Extended Data Figure~\ref{ExtFig4} displays the dependence of the temperature oscillation mean value $\langle T_1 \rangle$ on $J_{\rm in}$ as measured by the thermometer P$_3$ (at a distance $l_3\simeq 2.5$ \textmu m from the junction 2) and by P$_4$ (at a distance $l_4\simeq 5.7$ \textmu m from the junction 2) at a bath temperature of 25 mK. The experimental data immediately show that for $\langle T_1 \rangle \gtrsim 0.7$ K the electronic temperature of S$_1$ depends on the position along the electrode.

\begin{figure*}[h!]
\centering
%\hspace*{-2.em}
\includegraphics[width=0.6\textwidth]{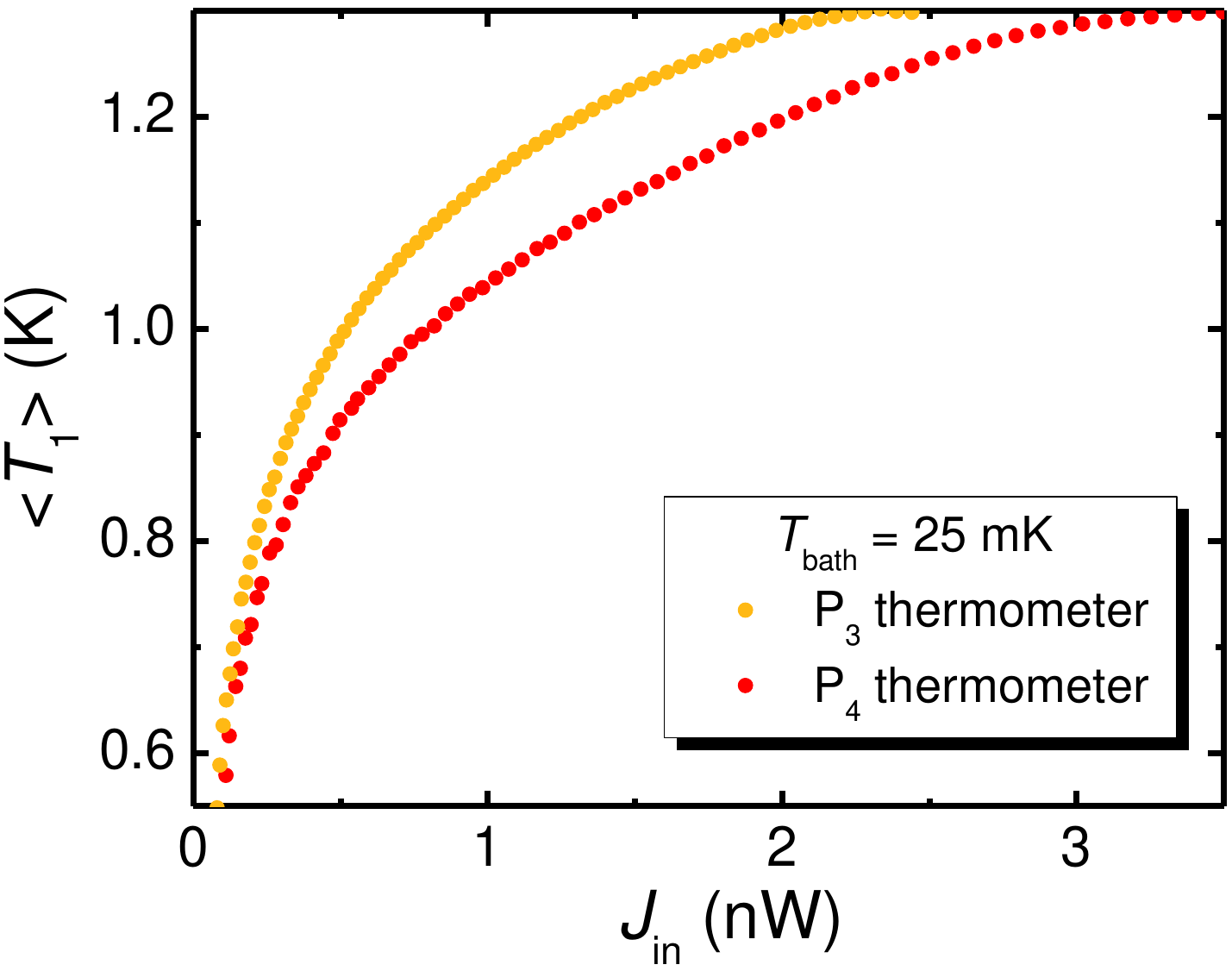}
%\vspace*{-4.ex}
\caption{\textbf{Temperature gradient along the electrode S$_1$.} Average electronic temperature $\langle T_1 \rangle$ vs. the injected Joule power $J_{\rm in}$ measured by the thermometers P$_3$ and P$_4$ at a bath temperature of 25 mK.\label{ExtFig4}} 
%% \vspace*{-3.ex}
\end{figure*}

As already mentioned in the Main Text, we model the S$_1$ island as a one-dimensional diffusive superconductor and its temperature profile $T_1(x)$ is given by the stationary heat diffusion equation~\cite{Timofeev2}:
\begin{equation}
\frac{d}{dx}\left\lbrace \kappa[T_1(x)] \frac{dT_1(x)}{dx}\right\rbrace=\frac{J_{\rm e-ph}(x)}{\mathcal{V}_1}+\frac{J_{\rm a}(x)}{\mathcal{V}_{\rm a}}+\sum_{i=3,4,5} \frac{J_{\rm th}^{\rm i}(x)}{\mathcal{V}_{\rm th}^{\rm i}}, \label{heateq2}
\end{equation}
where $\kappa[T_1(x)]=\sigma_{\rm N}/[2e^2k_{\rm B}T_1^2(x)]\int \mathrm{d}\epsilon \epsilon^2 \mathcal{F}[\epsilon, T_1(x)] \mathrm{sech}^2[\epsilon/2k_{\rm B}T_1(x)]$ is the electronic heat conductivity~\cite{Timofeev2} of S$_1$, $\mathcal{F}[\epsilon, T_1(x)]=\mathrm{cos}^2\lbrace \Im[t(x)]\rbrace$, $t(x)=\mathrm{arctanh}\lbrace \Delta_1[T_1(x)]/(\epsilon+i\Gamma_1)\rbrace$, $\sigma_{\rm N}=e^2 N(E_{\rm F}) D$ is the electrical conductivity in the normal state, $N(E_{\rm F})\sim 1.6\times 10^{47}$ m$^{-3}$J$^{-1}$ is the electronic density of states of Al calculated at the Fermi energy and $D\simeq 20$ cm$^{2}$s$^{-1}$ is the diffusion constant as measured in independent experiments on similar structures. Moreover, $\mathcal{V}_{\rm th}^{\rm i}=l_{\rm i}\, A$ and $\mathcal{V}_{\rm a}=l_{\rm a}\, A$, being $A$ the cross-section of S$_1$, $l_{\rm i}$ the length of the $i$-th thermometer junction (with $i=3,4,5$) and $l_{\rm a}$ the length of the SQUID junction "a". As boundary conditions, we impose $J_{\rm in} = -\kappa[T_1(0)]AT'_1(0)$ and $J_{\rm S_1S_2} = -\kappa[T_1(l)]AT'_1(l)$, determining the relation between the temperature gradients $T'_1(x)$ and the heat fluxes at the ends of S$_1$. Here, $x=0$ is set in correspondence to the heater junction 2, while $x=l=17.5$ \textmu m corresponds to the junction "j" of the SQUID.

Since the heater voltages in the experiment are $V_{\rm h}\gg \Delta_3(0)/e$ (we reach $V_{\rm h}\sim 40$ mV), the Joule power injected in the superconducting island is given by $I_{\rm h}V_{\rm h}/2$~\cite{GiazottoRev,Timofeev1}, where $I_{\rm h}$ is the measured current flowing through the heater junctions 1 and 2. On the other hand, as expressed in Eq.~\ref{Jtot} of the Main Text, $J_{\rm S_1S_2}$ is the electronic stationary heat current flowing through the temperature-biased JJ "j" of the SQUID~\cite{MakiGriffin,Guttman,GiazottoAPL}:
\begin{equation}
J_{\rm S_1S_2}(T_1,T_{\rm bath},\varphi_{\rm j})=J_{\rm qp}(T_1,T_{\rm bath})-J_{\rm int}(T_1,T_{\rm bath})\rm\; cos \varphi_{\rm j},\label{Jtot2}
\end{equation}
where $J_{\rm qp}(T_{1},T_{\rm bath})=(2/e^{2}R_{\rm j}) \displaystyle\int_{0}^{\infty} \mathrm{d}\epsilon \, \epsilon \mathcal{N}_1(\epsilon , T_{\rm 1})\mathcal{N}_2(\epsilon , T_{\rm bath}) \allowbreak [f(\epsilon ,T_{\rm 1})-f(\epsilon , T_{\rm bath})]$ is the incoherent term of the heat current, while $J_{\rm int}(T_{1},T_{\rm bath})=(2/e^{2}R_{\rm j}) \displaystyle\int_{0}^{\infty} \mathrm{d}\epsilon \, \epsilon \mathcal{M}_1(\epsilon , T_{\rm 1})\mathcal{M}_2(\epsilon , T_{\rm bath}) \allowbreak [f(\epsilon ,T_{\rm 1})-f(\epsilon , T_{\rm bath})]$ represents the thermal counterpart of the "quasiparticle-pair interference" component of the charge current tunneling through a JJ~\cite{Barone,Pop}. Here, $\mathcal{M}_{\rm n}(\epsilon,T) =\allowbreak |\Im[-i \Delta_{\rm n}(T)/\sqrt{(\epsilon\allowbreak +i\Gamma_{\rm n})^2-\Delta_{\rm n}^2(T)}]|$ is the Cooper pair BCS DOS in the n-th superconductor, with $\text{n}=1,2$.

In Eq.~\ref{heateq2} $J_{\rm a}=J_{\rm S_1S_3}(T_1,T_{\rm bath},\varphi_{\rm a})g_{\rm a}(x)$ takes in to account the energy released by S$_1$ to P$_6$ through the junction "a" of the SQUID, which is spatially limited by the broadened (with $\delta x =0.1$ \textmu m) step function $g_{\rm a}(x)= \lbrace 1/1+\text{exp}[(x-x_{\rm a,end})/\delta x]\rbrace \lbrace \text{exp}[(x-x_{\rm a,begin})/\delta x]/1+\text{exp}[(x-x_{\rm a,begin})/\delta x] \rbrace$. In the same way, the terms $J_{\rm th}^{\rm i}(x)=J_{\rm S_1S_3}(T_1,T_{\rm bath},0)g_{\rm i}(x)$ describe the heat currents flowing through the JJ of the thermometer P$_i$, with $i=3,4,5$. We note that $\delta x$ has been used for numerical reasons and does not affect the results provided that $\delta x$ is much smaller than the junction length ($x_{\rm a,end}-x_{\rm a,begin}\simeq 2$ \textmu m or $x_{\rm i,end}-x_{\rm i,begin}\geq 2$ \textmu m) and the distance between the junctions, which is of at least 0.5 \textmu m. Finally, $J_{\rm e-ph}[T_1(x)]$ represents the energy released by S$_1$ through the electron-phonon coupling, which reads~\cite{Timofeev1}:
\begin{align}
J_{\rm{e-ph}}[T_1(x),T_{\rm bath}]=&-\frac{\Sigma \mathcal{V_1}}{96 \zeta (5)k_{\rm B}^5}\int_{-\infty}^{\infty} \mathrm{d}E E \int_{-\infty}^{\infty} \mathrm{d}\epsilon \epsilon^2 \mathrm{sgn}(\epsilon) L(E,E+\epsilon,T_1)\left\lbrace\mathrm{coth}\left(\frac{\mathrm{\epsilon}}{2k_{\rm B}T_{\rm bath}}\right)\right.\notag \\
&\times [f^{\rm (1)}(E,T_1)-f^{\rm (1)}(E+\epsilon,T_1)]\left. -f^{\rm (1)}(E,T_1) f^{\rm (1)}(E+\epsilon,T_1)+1 \vphantom{\frac{1}{2}} \right\rbrace.
\end{align}
Here, $\Sigma=0.2\times 10^9$ Wm$^{-3}$K$^{-5}$ is the Al electron-phonon coupling constant~\cite{GiazottoRev}, $f^{\rm (1)}=f(-E)-f(E)$ and $L(E,E',T_1)=\mathcal{N}(E,T_1)\mathcal{N}(E',T_1)[1-\Delta^2(T_1)/(EE')]$.

\begin{figure*}[h!]
\centering
%\hspace*{-2.em}
\includegraphics[width=0.6\textwidth]{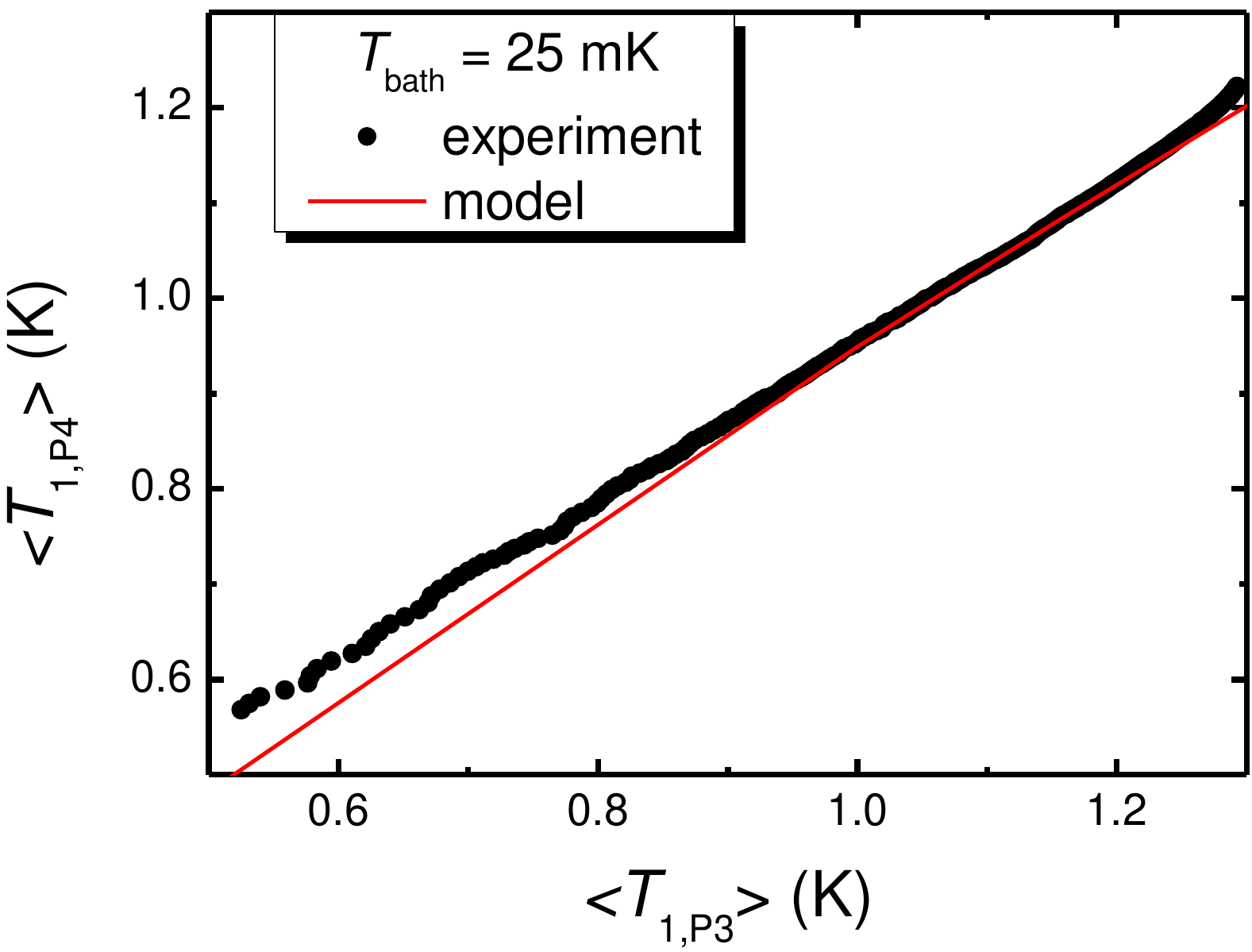}
%\vspace*{-4.ex}
\caption{\textbf{Theoretical model for the temperature gradient along S$_1$.} Average S$_1$ temperature $\langle T_{\rm 1,P_4} \rangle$ measured by the P$_4$ thermometer vs. the average temperature $\langle T_{\rm 1,P_3} \rangle$ measured by the thermometers P$_3$ at a bath temperature of 25 mK. The full circles are the experimental data, while the red solid line is the theoretical fit obtained from Eq.~\ref{heateq2}.
\label{ExtFig5}} 
%% \vspace*{-3.ex}
\end{figure*} 

The theoretical curves in Fig.~\ref{Fig3}a have been obtained by calculating $T_1(l_4)$ from Eq.~\ref{heateq2}, where $l_4=5.7$ \textmu m corresponds to the central position of the junction between S$_1$ and the P$_4$ thermometer. In Eq.~\ref{heateq2} we used the measured values of $R_{\rm 3}$, $R_{\rm 4}$, $R_{\rm 5}$, $R_{\rm a}$, $r_1$, $r_2$ as determined from the electrical characterization of the devices, while we set $\Gamma_1\simeq 0.1\Delta_1(0)$ accounting for an increased sub-gap transport. This value of $\gamma_1$ is necessary also at low temperatures and it might be caused by the voltage noise generated by the heater connections. The curves were fitted to the experimental data by inserting in the thermal model the flux-dependence of $\varphi_{\rm a}$ and $\varphi_{\rm j}$ calculated for the configuration in which only a circulating supercurrent is driven along the loop of the SQUID (see Fig.~\ref{Fig2}c of the Main Text). Furthermore, we varied $J_{\rm in}$ and $R_{\rm j}$ as the only fitting parameters. The injected Joule power $J_{\rm in}$ was reduced of about one order of magnitude with respect to the experimental values. This discrepancy might be ascribed to a \textit{local} non-equilibrium distribution of quasiparticles in proximity of the heaters junction~\cite{Kopnin}. Indeed, for $V_{\rm h}\gtrsim 100 k_{\rm B}T_{\rm c,1}$ the electron-electron and electron-phonon relaxation rates may become comparable, leading to the appearance of power-law tails in the electron distribution function that generate the emission of high-energy phonons~\cite{Kopnin}. The latter would carry away most of the energy provided by the current injected by the heaters, reducing the effective Joule power deposited in the island. Nevertheless, the effect results to be localized, as demonstrated by the good agreement between the experimental curves and the model based on quasi-equilibrium electronic distributions, as shown in Fig.~\ref{Fig3}a. On the other hand, we note that the value of $R_{\rm j}$ was varied from 25\% to 100\% of its nominal value, accounting for possible non-idealities of the AlO$_{\rm x}$ tunnel junction that can induce deviations from Eq.~\ref{Jtot2}.

The validity of our thermal model is confirmed by Extended Data Figure~\ref{ExtFig5}, which shows the average S$_1$ temperature measured by the P$_4$ thermometer as a function of that obtained with P$_3$, together with the theoretical fit calculated from Eq.~\ref{heateq2}. The good agreement between the experiment and the model is obtained by using the structural parameters obtained from the electrical measurements with the exception of $\gamma_1$ and $J_{\rm in}$ that were treated as explained in the previous paragraph. 
%Even though the $T_{\rm 1,P_4}$ vs. $T_{\rm 1,P_3}$ curve does not depend strongly on the value of $\gamma_1$, its value is necessary to fit the temperature oscillations shown in Fig.~\ref{Fig3}a. 
%On the other hand, $J_{\rm in}$} was reduced of about one order of magnitude with respect to the experimental values. This discrepancy might be ascribed to a \textit{local} non-equilibrium distribution of quasiparticles in proximity of the heaters junction~\cite{Kopnin}. Indeed, for $V_{\rm h}\gtrsim 100 k_{\rm B}T_{\rm c,1}$ the electron-electron and electron-phonon relaxation rates may become comparable, leading to the appearance of power-law tails in the electron distribution function that generate the emission of high-energy phonons~\cite{Kopnin}. The latter would carry away most of the energy provided by the current injected by the heaters, reducing the effective Joule power deposited in the island. Nevertheless, the effect results to be localized, as demonstrated by the good agreement between the experimental temperature dependence and the model based on quasi-equilibrium electronic distributions, as shown in Extended Data Fig.~\ref{ExtFig5}. %Indeed, the spatial range of the non-equilibrium distribution is limited by the electron-electron relaxation length $\sqrt{D/\gamma_{\rm e-e}}\sim 1-4$ \textmu m, where $\gamma\sim 10^8-10^9$ s$^{-1}$ is the electron-electron relaxation rate of Al at $T_1\sim T_{\rm c,1}$~\cite{GiazottoRev,Timofeev1}. 

Finally, the power generated by the electron-photon coupling between the two branches of the SQUID~\cite{Bosisio} has been estimated to be about two orders of magnitude smaller than the contribution due to quasiparticle tunneling and electron-phonon coupling, thereby leading to insignificant corrections of our results.

%%%%%%%%%%%%%%%%%%%%%%%%%%%%%%%%%%%%%%%%%%%%%%%%%%%%%%%%%%%%
%%%%%%%%%%%%%%%%%%%%%%%%%%%%%%%%%%%%%%%%%%%%%%%%%%%%%%%%%%%%
%%%%%%%%%%%%%%%%%%%%%%%%%%%%%%%%%%%%%%%%%%%%%%%%%%%%%%%%%%%%
%%%%%%%%%%%%%%%%%%%%%%%%%%%%%%%%%%%%%%%%%%%%%%%%%%%%%%%%%%%%
%%%%%%%%%%%%%%%%%%%%%%%%%%%%%%%%%%%%%%%%%%%%%%%%%%%%%%%%%%%%
%%%%%%%%%%%%%%%%%%%%%%%%%%%%%%%%%%%%%%%%%%%%%%%%%%%%%%%%%%%%
%FIGURES
%%%%%%%%%%%%%%%%%%%%%%%%%%%%%%%%%%%%%%%%%%%%%%%%%%%%%%%%%%%%


\section*{References}
\begin{thebibliography}{99}
\bibitem{Josephson} Josephson, B. D. Possible new effects in superconductive tunneling. \textit{Phys. Lett.} \textbf{1}, 251-253 (1962).
\bibitem{MakiGriffin} Maki, K. \& Griffin, A. Entropy transport between two superconductors by electron tunneling. \textit{Phys. Rev. Lett.} \textbf{15}, 921-923 (1965).
\bibitem{GiazottoNature} Giazotto, F. \& Mart\'inez-P\'erez, M. J. The Josephson heat interferometer. \textit{Nature} \textbf{492}, 401-405 (2012).
\bibitem{MartinezNature} Mart\'inez-P\'erez, M. J. \& Giazotto, F. A quantum diffractor for thermal flux. \textit{Nat. Commun.} \textbf{5}, 3579 (2014).
\bibitem{Ryazanov} Ryazanov, V. V. \textit{et al.} Coupling of two superconductors through a ferromagnet: evidence for a $\pi$ junction. \textit{Phys. Rev. Lett.} \textbf{86}, 2427-2430 (2001).
%\bibitem{Kontos} Kontos, T. \textit{et al.} Josephson junction through a thin ferromagnetic layer: negative coupling. \textit{Phys. Rev. Lett.} \textbf{89}, 137007 (2002).
\bibitem{Gingrich} Gingrich, E. C. \textit{et al.} Controllable $0-\pi$ Josephson junctions containing a ferromagnetic spin valve. \textit{Nature Phys.} doi:10.1038/nphys3681 (2016).
\bibitem{Baselmans} Baselmans, J. J. A., Morpurgo, A. F., van Wees, B. J. \& Klapwijk, T. M. Reversing the direction of the supercurrent in a controllable Josephson junction. \textit{Nature} \textbf{397}, 43-45 (1999).
\bibitem{Feofanov} Feofanov, A. K. \textit{et al.} Implementation of superconductor/ferromagnet/superconductor $\pi$-shifters in superconducting digital and quantum circuits. \textit{Nature Phys.} \textbf{6}, 593-597 (2010).
\bibitem{Holmes} Holmes, D. S., Ripple, A. L. \& Manheimer, M. A. Energy-efficient superconducting computing-power budgets and requirements. \textit{IEEE Trans. Appl. Supercond.} \textbf{23}, 170610 (2013).
\bibitem{FornieriPRB} Fornieri, A., Timossi, G., Bosisio, R., Solinas, P. \& Giazotto, F. Negative differential thermal conductance and heat amplification in superconducting hybrid devices. \textit{Phys. Rev. B} \textbf{93}, 134508 (2016).
\bibitem{MartinezRev} Mart\'inez-P\'erez, M. J., Solinas, P. \& Giazotto, F. Coherent caloritronics in Josephson-based nanocircuits. \textit{J. Low Temp. Phys.} \textbf{175}, 813-837 (2014).
\bibitem{FornieriNature} Fornieri, A., Blanc, C., Bosisio, R., D'Ambrosio, S. \& Giazotto, F. Nanoscale phase engineering of thermal transport with a Josephson heat modulator. \textit{Nat. Nanotechnol.} \textbf{11}, 258-262 (2016).
\bibitem{MartinezNatRect} Mart\'inez-P\'erez, M. J., Fornieri, A. \& Giazotto, F. Rectification of electronic heat current by a hybrid thermal diode. \textit{Nat. Nanotechnol.} \textbf{10}, 303-307 (2015).
\bibitem{MartinezAPL} Mart\'inez-P\'erez, M. J. \& Giazotto, F. Efficient phase-tunable Josephson thermal rectifier. \textit{Appl. Phys. Lett.} \textbf{102}, 182602 (2013).
\bibitem{Tinkham} Tinkham, M. \textit{Introduction to Superconductivity} (McGraw-Hill, 1996) and references therein.
\bibitem{Guttman} Guttman, G. D., Nathanson, B., Ben-Jacob, E. \& Bergman, D. J. Phase-dependent thermal transport in Josephson junctions. \textit{Phys. Rev. B} \textbf{55}, 3849-3855 (1997).
\bibitem{GiazottoAPL} Giazotto, F. \& Mart\'inez-P\'erez, M. J. Phase-controlled superconducting heat-flux quantum modulator. \textit{Appl. Phys. Lett.} \textbf{101}, 102601 (2012).
\bibitem{Barone} A. Barone and G. Paternò, \textit{Physics and Applications of the Josephson Effect} (Wiley, New York, 1982).
\bibitem{Pop} I. M. Pop \textit{et al.} Coherent suppression of electromagnetic dissipation due to superconducting quasiparticles. \textit{Nature} \textbf{508}, 369-372 (2014).
\bibitem{GiazottoRev} Giazotto, F., Heikkil\"{a}, T. T., Luukanen, A., Savin, A. M. \& Pekola, J. P. Opportunities for mesoscopics in thermometry and refrigeration: Physics and applications. \textit{Rev. Mod. Phys.} \textbf{78}, 217-274 (2006).
\bibitem{Tirelli} Tirelli, S. \textit{et al.}. Manipulation and generation of supercurrent in out-of-equilibrium Josephson tunnel nanojunctions. Phys. Rev. Lett. \textbf{101}, 077004 (2008).
\bibitem{Quaranta} Quaranta, O., Spathis, P., Beltram, F. \& Giazotto, F. Cooling electrons from 1 to 0.4 K with V-based nanorefrigerators. Appl. Phys. Lett. \textbf{98}, 032501 (2011).
\bibitem{SQUIDhandbook} Clarke, J. \& Braginski, A. I. (eds) \textit{The SQUID Handbook} (Wiley-VCH, 2004).
\bibitem{Wellstood} Wellstood, F. C., Urbina, C. \& Clarke, J. Hot-electron effects in metals. \textit{Phys. Rev. B} \textbf{49}, 5942-5955 (1994).
\bibitem{Timofeev1} Timofeev, A. V. \textit{et al}. Recombination-limited energy relaxation in a Bardeen-Cooper-Schrieffer superconductor.
\textit{Phys. Rev. Lett.} \textbf{102}, 017003 (2009).
\bibitem{Timofeev2} Timofeev, A. V., Helle, M., Meschke, M., M\"{o}tt\"{o}nen, M. \& Pekola, J. P. Electronic refrigeration at the quantum limit. \textit{Phys. Rev. Lett.} \textbf{102}, 200801 (2009).
%\bibitem{Bardeen} Bardeen, J., Rickayzen, G. \& Tewordt, L. Theory of the thermal conductivity of superconductors. \textit{Phys. Rev.} \textbf{113}, 982-994 (1959).
\bibitem{Pascal} Pascal, L. M. A., Courtois, H. \& Hekking, F. W. J. Circuit approach to photonic heat transport. \textit{Phys. Rev, B} \textbf{83}, 125113 (2011).
\bibitem{meschke} Meschke, M., Guichard, W. \& Pekola, J. P. Single-mode heat conduction by photons. \textit{Nature} \textbf{444}, 187-190 (2006).
%\bibitem{schmidt} Schmidt, D. R., Schoelkopf, R. J. \& Cleland, A. N. Photon-mediated thermal relaxation of electrons in nanostructures. \textit{Phys. Rev. Lett.} \textbf{93}, 045901 (2004).
\bibitem{Bosisio} Bosisio, R., Solinas, P., Braggio, A. \& Giazotto, F. Photonic heat conduction in Josephson-coupled Bardeen-Cooper-Schrieffer superconductors. \textit{Phys. Rev. B} \textbf{93}, 144512 (2016).
%\bibitem{LiWangCasatiAPL} Li, B., Wang, L. \& Casati, G. Negative differential thermal resistance and thermal transistor. Appl. Phys. Lett. \textbf{88}, 143501 (2006).
\bibitem{LiRev} Li, N. \textit{et al}. Phononics: manipulating heat flow with electronic analogs and beyond. \textit{Rev. Mod. Phys.} \textbf{84}, 1045-1066 (2012).


\section*{Methods References}
\bibitem{Dynes} R. C. Dynes, V. Narayanamurty \& J. P. Garno. Direct measurement of quasiparticle-lifetime broadening in a strong-coupled superconductor. \textit{Phys. Rev. Lett.} \textbf{41}, 1509-1512 (1978).
\bibitem{PekolaPRL2010} J. P. Pekola, V. F. Maisi, S. Kafanov, N. Chekurov, A. Kemppinen, Yu. A. Pashkin, O.-P. Saira, M. M\"ott\"onen, and J. S. Tsai, Phys. Rev. Lett. \textbf{105}, 026803 (2010).
\bibitem{GiazottoJap} Giazotto, F. \& Pekola, J. P. Josephson tunnel junction controlled by quasiparticle injection. J. Appl. Phys. \textbf{97}, 023908 (2005).
\bibitem{Kopnin}  Kopnin, N. B., Galperin, Y. M., Bergli, J. \& Vinokur, V. M. Nonequilibrium electrons in tunnel structures under high-voltage injection. \textit{Phys. Rev. B} \textbf{80}, 134502 (2009).
\end{widetext}
\end{thebibliography}
\end{document}